\documentclass[aps,preprint]{revtex4}%
\usepackage{amsfonts}
\usepackage{amsmath}
\usepackage{amssymb}
\usepackage{graphicx}%
\providecommand{\U}[1]{\protect\rule{.1in}{.1in}}
\begin{document}
\preprint{HEP/123-qed}
\title[Short title for running header]{Quantum particle on a M\"{o}bius strip, coherent states and projection
operators }
\author{Diego Julio Cirilo-Lombardo}
\affiliation{Bogoliubov Laboratory of Theoretical Physics, Joint Institute for Nuclear
Research, 141980 Dubna, Moscow region, Russian Federation}
\affiliation{International Institute of Physics,Capim Macio - 59078-400 - Natal-RN, Brazil}
\author{}
\affiliation{}
\author{}
\affiliation{}

\begin{abstract}
The coherent states for a quantum particle on a M\"{o}bius strip are
constructed and their relation with the natural phase space for fermionic
fields is shown. The explicit comparison of the obtained states with previous
works where the cylinder quantization was used and the spin 1/2 was introduced
by hand is given, and the relation between the geometrical phase space,
constraints and projection operators is analyzed and discussed.

\end{abstract}
\volumeyear{year}
\volumenumber{number}
\issuenumber{number}
\eid{identifier}
\date[Date text]{date}
\received[Received text]{date}

\revised[Revised text]{date}

\accepted[Accepted text]{date}

\published[Published text]{date}

\startpage{101}
\endpage{102}
\maketitle
\tableofcontents

\section{Introduction}

Coherent States have attracted much attention in many branches of physics [1].
In spite of their importance, the theory of CS when the configuration space
has non trivial topology is far from complete. CS for a quantum particle on a
circle [3] and a sphere have been introduced very recently, and also in the
case of torus [4,7]. If well , in all these works the different constructions
of the CS for the boson case is practically straighforward, the simple
addition by hand of 1/2 to the angular momentum operator $J$ for the fermionic
case into the corresponding CS remains obscure and non-natural. The question
that naturally arises is: there exists any geometry for the phase space in
which the CS construction leads precisely a fermionic quantization condition?
Recently in previous works [9] we demonstrate the positive answer to this
question showing that the CS for a quantum particle on the M\"{o}bius strip
geometry is the natural candidate to describe fermions exactly as the cylinder
geometry for bosons. Then, the purpose of this paper is to analyze deeply this
relation between the coherent states and the geometry of the physical phase
space taking into account two important roles playing by the CS: as projector
operators [10] and as the main link between classical and quantum formulations
of a given system [5].

\section{Requirements for the Coherent states}

It is well know that coherent states provide naturally a close connection
between classical and quantum formulations of a given system. A suitable set
of requirements for these states is given, in association with a specific
Hamiltonian operator $\mathcal{H}$, by

(a) Continuity: $\left(  J^{\prime},\gamma^{\prime}\right)  \rightarrow\left(
J,\gamma\right)  \Rightarrow\left\vert J^{\prime},\gamma^{\prime}\right\rangle
\rightarrow\left\vert J,\gamma\right\rangle .$

(b) Resolution of the unity: $\mathbb{I=}\int\left\vert J,\gamma\right\rangle
\left\langle J,\gamma\right\vert d\mu\left(  J,\gamma\right)  .$

(c) Temporal stability: $e^{-i\mathcal{H}t}\left\vert J,\gamma\right\rangle
=\left\vert J,\gamma+\omega t\right\rangle ,\omega=constant.$

(d) Action identity: $\left\langle J,\gamma\right\vert \mathcal{H}\left\vert
J,\gamma\right\rangle =\omega J.$

The first two requirements emphasize the fact that the identity operator may
be understood in a restricted sense, namely as a projector onto a finite or
infinite subspace. The third requirement ensures that the time evolution of
any coherent state is always a coherent state. As was showed clearly by Gazeau
and Klauder in [4], in this evolution, $J$ remains constant while $\gamma$
increases linearly. These properties are similar to the classical behavior of
action-angle variables. If $J$ and $\gamma$ denote canonical action-angle
variables, they would enter the classical action in the following form%
\[
I=\int_{0}^{T}\left(  J\overset{.}{\gamma}-\omega J\right)  dt.
\]
As is easily seen, the classical action can be viewed as the restricted
evaluation of the quantum action functional:%
\[
I=\int_{0}^{T}\left[  i\left\langle J,\gamma\right\vert \frac{d}{dt}\left\vert
J,\gamma\right\rangle -\left\langle J,\gamma\right\vert \mathcal{H}\left\vert
J,\gamma\right\rangle \right]  dt.
\]
for different paths $\{\left\vert J\left(  t\right)  ,\gamma\left(  t\right)
\right\rangle :0\leq t\leq T\}$ lying in a two dimensional manifold in Hilbert
space. Thus the fourth requirement simply codifies the fact that the two
coordinates $\left(  J,\gamma\right)  $ are canonical action-angle variables
(it will follow that the kinematical term is $J\overset{.}{\gamma}$ as needed
[5]). At this point seems to be necessary to make the following observations:
firstly, the physical meaning of the third requirement is to assert that the
path in the Hilbert space represented by $\{\left\vert J,\gamma+\omega
t\right\rangle :0\leq t\leq T\}$ is actually the true quantum temporal for the
quantum Hamiltonian $\mathcal{H}$. Then, the restricted quantum action
functional in this case is \textit{exact}, see [1] - Gazeau and references
therein; a wider set of variatonal paths starting at $\left\vert
J,\gamma\right\rangle $ at $t=0$ leads to the same extreme path. Secondly, it
is well known that the lack of uniqueness in the possible families of CS
corresponding to a given Hamiltonian with discrete spectrum is because the
fourth requirement was not taken into account\footnotemark

\section{Geometry of the M\"{o}bius band and dynamics}

The position of a point into the M\"{o}bius strip geometry can be
parameterized as%

\begin{equation}%
\begin{tabular}
[c]{lll}%
$P_{0}=\left(  X_{0},Y_{0},Z_{0}\right)  $ & $,$ & $P_{1}=\left(  X_{0}%
+X_{1},Y_{0}+Y_{1},Z_{0}+Z_{1}\right)  $%
\end{tabular}
\ \ \tag{1}%
\end{equation}

The coordinates of $P_{0}$ describes the central cylinder (generated by the
invariant fiber of the middle of the weight of the strip)%
\begin{equation}%
\begin{tabular}
[c]{lll}%
$Z_{0}=l\ ,\ \ \ $ & $X_{0}=R\ Cos\varphi,\ \ \ \ $ & $Y_{0}=R\ Sin\varphi,$%
\end{tabular}
\ \ \ \ \ \tag{2}%
\end{equation}
(this is topological invariant of the geometry under study).

The coordinates of $P_{1}$ (the boundaries of the M\"{o}bius band) are of
$P_{0}$ (the cylinder) plus
\begin{equation}%
\begin{tabular}
[c]{lll}%
$Z_{1}=r\ Cos\theta,\ \ \ \ \ $ & $X_{1}=r\ Sin\theta\ Cos\varphi\ ,\ \ \ $ &
$Y_{1}=r\ Sin\theta\ Sin\varphi,$%
\end{tabular}
\ \ \ \ \ \tag{3}%
\end{equation}
The weight of the band is obviously $2r$, then our space of phase is embedded
into of the Torus%
\begin{equation}%
\begin{tabular}
[c]{l}%
$X=R\ Cos\varphi+r\ Sin\theta\ Cos\varphi$\\
$Y=R\ Sin\varphi+r\ Sin\theta\ Sin\varphi$\\
$Z=l+r\ Cos\theta$%
\end{tabular}
\ \ \ \ \ \tag{4}%
\end{equation}
The important point is that the angles are not independent in the case of the
M\"{o}bius band and are related by the following constraint%
\begin{equation}
\theta=\frac{\varphi+\pi}{2} \tag{5}%
\end{equation}
It is very important, this constraint effectively reduces the degree of
freedom from the torus to the unnoriented surface.

In order to study the dynamics in this non-trivial geometry, we construct the
non-relativistic Lagrangian%
\begin{equation}
L=\frac{1}{2}m(\overset{.}{X}^{2}+\overset{.}{Y}^{2}+\overset{.}{Z}^{2})
\tag{6}%
\end{equation}%
\begin{equation}
L=\frac{1}{2}\left\{  \overset{.}{\varphi}^{2}\left[  (1+r\ Cos\left(
\varphi/2\right)  \ )^{2}+\frac{r^{2}}{4}\right]  -r\ Cos\left(
\varphi/2\right)  \ \overset{.}{Z}_{0}\overset{.}{\varphi}+\left(  \overset
{.}{Z}_{0}\right)  ^{2}\right\}  \tag{7}%
\end{equation}
From the above expression the equations of motion are%
\begin{equation}
\frac{\partial L}{\partial\overset{.}{\varphi}}=\overset{.}{\varphi}\left[
(1+r\ Cos\left(  \varphi/2\right)  \ )^{2}+\frac{r^{2}}{4}\right]  -\frac
{r\ }{2}Cos\left(  \varphi/2\right)  \ \overset{.}{Z}_{0} \tag{8}%
\end{equation}%
\begin{equation}
\frac{\partial L}{\partial\overset{.}{Z}_{0}}=-\frac{r\ }{2}Cos\left(
\varphi/2\right)  \ \overset{.}{\varphi}+\overset{.}{Z}_{0} \tag{9}%
\end{equation}%
\begin{equation}
\frac{\partial L}{\partial\varphi}=\overset{.}{\frac{r\ }{2}Sin\left(
\varphi/2\right)  \overset{.}{\varphi}}\left[  -\overset{.}{\varphi
}(1+r\ Cos\left(  \varphi/2\right)  \ )+\frac{\overset{.}{Z}}{2}\right]
\tag{10}%
\end{equation}%
\begin{equation}
\frac{\partial L}{\partial Z_{0}}=0 \tag{11}%
\end{equation}
Taking account that $Z_{0}$ is a cyclic coordinate, we have the following
constraint%
\begin{equation}
\left(  \frac{\partial L}{\partial\overset{.}{Z}_{0}}\right)  ^{.}%
-\frac{\partial L}{\partial Z_{0}}=0\Rightarrow\frac{\partial L}%
{\partial\overset{.}{Z}_{0}}=L_{0}=-\frac{r\ }{2}Cos\left(  \varphi/2\right)
\ \overset{.}{\varphi}+\overset{.}{Z}_{0} \tag{12}%
\end{equation}
then, looking for the dynamical expressions for $\overset{.}{\varphi}$
\begin{equation}
\frac{\partial L}{\partial\overset{.}{\varphi}}=\overset{.}{\varphi}\left[
(1+r\ Cos\left(  \varphi/2\right)  \ )^{2}+\frac{r^{2}}{4}Sin^{2}\left(
\varphi/2\right)  \right]  -\frac{r\ }{2}Cos\left(  \varphi/2\right)
\ L_{0}=J \tag{13}%
\end{equation}
From the Lagrangian (7) the Hamiltonian is not difficult to obtain%
\begin{align}
H  &  =p_{\varphi}\overset{.}{\varphi}+p_{z_{0}}\overset{.}{Z}_{0}-L\tag{14}\\
&  =\frac{1}{2}\left\{  \overset{.}{\varphi}^{2}\left[  (1+r\ Cos\left(
\varphi/2\right)  \ )^{2}+\frac{r^{2}}{4}\right]  -r\ Cos\left(
\varphi/2\right)  \ \overset{.}{Z}_{0}\overset{.}{\varphi}+\left(  \overset
{.}{Z}_{0}\right)  ^{2}\right\}  =L\nonumber
\end{align}
that trough the constraint (12) takes the most compact form%
\begin{equation}
H=\frac{1}{2}\left\{  \overset{.}{\varphi}^{2}\left[  (1+r\ Cos\left(
\varphi/2\right)  \ )^{2}-\frac{r^{2}}{4}Cos\varphi\right]  +L_{0}%
^{2}\right\}  \tag{15}%
\end{equation}
As usual in the Hamiltonian formulation, it is convenient to introduce
\begin{equation}
\mathbb{J}\equiv\overset{.}{\varphi}=\frac{\left(  J+\frac{rL_{0}Cos\left(
\varphi/2\right)  }{2}\right)  }{\left[  (1+r\ Cos\left(  \varphi/2\right)
\ )^{2}+\frac{r^{2}}{4}Sin^{2}\left(  \varphi/2\right)  \right]  } \tag{16}%
\end{equation}
then, finally the expression (15) takes the form%
\begin{align}
H  &  =\left\{  \frac{\left(  \widehat{J}+\frac{rL_{0}Cos\left(
\varphi/2\right)  }{2}\right)  ^{2}\left[  (1+r\ Cos\left(  \varphi/2\right)
\ )^{2}-\frac{r^{2}}{4}Cos\varphi\right]  }{\left[  (1+r\ Cos\left(
\varphi/2\right)  \ )^{2}+\frac{r^{2}}{4}Sin^{2}\left(  \varphi/2\right)
\right]  ^{2}}+L_{0}^{2}\right\} \tag{17}\\
&  =\frac{1}{2}\left\{  \mathbb{J}^{2}\left[  (1+r\ Cos\left(  \varphi
/2\right)  \ )^{2}-\frac{r^{2}}{4}Cos\varphi\right]  +L_{0}^{2}\right\}
\nonumber
\end{align}
These expressions above involving geometry and dynamics on the M\"{o}bius
strip (MS) will be utilized at the quantum level in next Sections.

\section{Abstract coherent states}

\bigskip In order to introduce the coherent states for a quantum particle on
the M\"{o}bius strip geometry we follow the Barut-Girardello construction [2]
and we seek the CS as the solution of the eigenvalue equation%

\begin{equation}
X\left\vert \xi\right\rangle =\xi\left\vert \xi\right\rangle \tag{18}%
\end{equation}
with complex $\xi$. Similarly as the standard case where the coherent states
$\left\vert z\right\rangle $ satisfy the eigenvalue equation where
$z\in\mathbb{C}$:%

\begin{equation}
e^{ia}\left\vert z\right\rangle =e^{iz}\left\vert z\right\rangle \tag{19}%
\end{equation}
where $a$ is the standard bosonic annihilation operator with $\widehat{q}$ and
$\widehat{p}$ the position and momentum operators respectively, then we can define%

\begin{equation}
X:=e^{i(\widehat{\varphi}+i\widehat{J})}\tag{20}%
\end{equation}
Taking $R=1$ and inserting (5) into (4) we obtain the parametrization of the band%

\begin{equation}%
\begin{tabular}
[c]{l}%
$X=\ Cos\varphi+r\ Cos\left(  \varphi/2\right)  \ Cos\varphi$\\
$Y=Sin\varphi+r\ Cos\left(  \varphi/2\right)  \ Sin\varphi$\\
$Z=l+r\ Sin\left(  \varphi/2\right)  $%
\end{tabular}
\ \ \ \tag{21}%
\end{equation}
Taking into account on the initial condition, and the transformations%

\begin{equation}%
\begin{tabular}
[c]{l}%
$X^{\prime}=e^{-Z}X$\\
$Y^{\prime}=e^{-Z}Y$\\
$Z^{\prime}=Z$%
\end{tabular}
\ \ \ \ \ \tag{22}%
\end{equation}
we finally have%
\begin{equation}
\xi=e^{-\left(  l+r\ Sin\left(  \varphi/2\right)  \right)  +i\varphi}\left(
1+r\ Cos\left(  \varphi/2\right)  \right)  \tag{23}%
\end{equation}
Inserting above expression in the expansion of the coherent state in the $j$
basis we obtain the CS\ in explicit form%
\begin{align}
\left\vert \xi\right\rangle  &  =\overset{\infty}{\underset{j=-\infty}{\sum}%
}\xi^{-j}e^{-\frac{j^{2}}{2}}\left\vert j\right\rangle \tag{24}\\
&  =\overset{\infty}{\underset{j=-\infty}{\sum}}e^{\left[  \left(
l+r\ Sin\left(  \varphi/2\right)  \right)  -ln\left(  1+r\ Cos\left(
\varphi/2\right)  \right)  -i\varphi\right]  j}e^{-\frac{j^{2}}{2}}\left\vert
j\right\rangle \nonumber\\
&  =\overset{\infty}{\underset{j=-\infty}{\sum}}e^{l^{\prime}j-i\varphi
j}e^{-\frac{j^{2}}{2}}\left\vert j\right\rangle
\ \ \ \ \ \ \ \ \ \ \ \nonumber
\end{align}
From (24), the fiducial vector is%
\begin{equation}
\left\vert 1\right\rangle =\overset{\infty}{\underset{j=-\infty}{\sum}%
}e^{-\frac{j^{2}}{2}}\left\vert j\right\rangle \tag{25}%
\end{equation}
then%
\begin{equation}
\left\vert \xi\right\rangle =e^{-\left(  ln\xi\right)  \widehat{J}}\left\vert
1\right\rangle \tag{26}%
\end{equation}
$($ in the expression (25) the sum absolutely converges to a finite value
($\Theta_{3}\left(  0\mid e^{-1/2}\right)  $ ) for $j\in\mathbb{R}$ .). As it
is easily seen the fiducial vector $\left\vert 1\right\rangle =\left\vert
0,0\right\rangle _{r=0}$ in the $\left(  l,\varphi\right)  $ parametrization,
and this fact permits us rewrite expression (26) as%
\begin{equation}
\left\vert l,\varphi\right\rangle =e^{\left[  \left(  l+r\ Sin\left(
\varphi/2\right)  \right)  -ln\left(  1+r\ Cos\left(  \varphi/2\right)
\right)  -i\varphi\right]  j}\left\vert 0,0\right\rangle _{r=0}\tag{27}%
\end{equation}
\bigskip The apparent singularity in (24) corresponding to the case $\xi=0$
are only for assymptotic values of $\left(  l+r\ Sin\left(  \varphi/2\right)
\right)  .$ Notice that in (23) the quantity $\left(  1+r\ Cos\left(
\varphi/2\right)  \right)  $ never is zero due that $0<r<R$ with $R=1.$ The
overlapping and non-ortogonality formulas are explicitly derived from (26)
\begin{equation}
\left\langle \xi\right.  \left\vert \eta\right\rangle =\overset{\infty
}{\underset{j=-\infty}{\sum}}\left(  \xi^{\ast}\eta\right)  ^{-j}e^{-j^{2}%
}=\Theta_{3}\left(  \frac{i}{2\pi}ln\left(  \xi^{\ast}\eta\right)  \mid
\frac{i}{\pi}\right)  \tag{28}%
\end{equation}
and
\begin{equation}
\left\langle l,\varphi\right.  \left\vert h,\psi\right\rangle =\Theta
_{3}\left(  \frac{i}{2\pi}\left(  \varphi-\psi\right)  -\frac{l^{\prime
}+h^{\prime}}{2}\frac{i}{\pi}\mid\frac{i}{\pi}\right)  ,\tag{29}%
\end{equation}
respectively.where we have been defined $l^{\prime}$ and $h^{\prime}$ in order
to have more compact expressions as follows:%
\begin{align*}
l^{\prime}  & \equiv\left(  l+r\ Sin\left(  \varphi/2\right)  \right)
-ln\left(  1+r\ Cos\left(  \varphi/2\right)  \right)  \\
h^{\prime}  & \equiv\left(  l+r\ Sin\left(  \psi/2\right)  \right)  -ln\left(
1+r\ Cos\left(  \psi/2\right)  \right)
\end{align*}
Finally, the normalization as a function of $\Theta_{3}$ yields%
\begin{equation}
\left\langle \xi\right.  \left\vert \xi\right\rangle =\Theta_{3}\left(
\frac{i}{\pi}ln\left\vert \xi\right\vert \mid\frac{i}{\pi}\right)  \tag{30}%
\end{equation}%
\begin{equation}
\left\langle l,\varphi\right.  \left\vert l,\varphi\right\rangle =\Theta
_{3}\left(  \frac{il^{\prime}}{\pi}\mid\frac{i}{\pi}\right)  \tag{31}%
\end{equation}

\section{The physical phase space and the natural quantization}

From equations
\begin{equation}
\widehat{J}\left\vert j\right\rangle =j\left\vert j\right\rangle \tag{32}%
\end{equation}%
\begin{equation}
\left\vert l,\varphi\right\rangle =\overset{\infty}{\underset{j=-\infty}{\sum
}}e^{l^{\prime}j-i\varphi j}e^{-\frac{j^{2}}{2}}\left\vert j\right\rangle
\ \ \tag{33}%
\end{equation}%
\begin{equation}
\left\langle j\right.  \left\vert l,\varphi\right\rangle =e^{l^{\prime
}j-i\varphi j}e^{-\frac{j^{2}}{2}} \tag{34}%
\end{equation}%
\begin{equation}
\left\langle l,\varphi\right.  \left\vert l,\varphi\right\rangle
=\overset{\infty}{\underset{j=-\infty}{\sum}}e^{l^{\prime}j}e^{-j^{2}}%
=\Theta_{3}\left(  \frac{il^{\prime}}{\pi}\mid\frac{i}{\pi}\right)  \tag{35}%
\end{equation}
we notice that the normalization, which for the cylinder (boson case) doesn't
depend on $\varphi$, depends now on $\varphi$ through $l^{\prime}\equiv\left(
l+r\ Sin\left(  \varphi/2\right)  \right)  -ln\left(  1+r\ Cos\left(
\varphi/2\right)  \right)  $. Also%
\begin{equation}
\widehat{J}\left\vert l,\varphi\right\rangle =\overset{\infty}{\underset
{j=-\infty}{\sum}}e^{l^{\prime}j-i\varphi j}e^{-\frac{j^{2}}{2}}j\left\vert
j\right\rangle \ \tag{36}%
\end{equation}
then%
\begin{equation}
\frac{\left\langle \xi\right\vert \widehat{J}\left\vert \xi\right\rangle
}{\left\langle \xi\right.  \left\vert \xi\right\rangle }=\frac{\left\langle
l,\varphi\right\vert \widehat{J}\left\vert l,\varphi\right\rangle
}{\left\langle l,\varphi\right.  \left\vert l,\varphi\right\rangle }=\frac
{1}{2\Theta_{3}\left(  \frac{il^{\prime}}{\pi}\mid\frac{i}{\pi}\right)  }%
\frac{\partial\Theta_{3}\left(  \frac{il^{\prime}}{\pi}\mid\frac{i}{\pi
}\right)  }{\partial l}. \tag{37}%
\end{equation}
Taking into account the identity%
\begin{equation}
\Theta_{3}\left(  \frac{il^{\prime}}{\pi}\mid\frac{i}{\pi}\right)  =e^{\left(
l^{\prime}\right)  ^{2}}\sqrt{\pi}\Theta_{3}\left(  l^{\prime}\mid
i\pi\right)  \tag{38}%
\end{equation}
coming from the general formula%
\begin{equation}
\Theta_{3}\left(  \frac{\nu}{\tau}\mid-\frac{1}{\tau}\right)  =e^{i\pi\nu
^{2}/\tau}\sqrt{\pi}\Theta_{3}\left(  \nu\mid\tau\right)  \tag{39}%
\end{equation}
we arrive at the following expression%
\begin{equation}
\frac{\left\langle \xi\right\vert \widehat{J}\left\vert \xi\right\rangle
}{\left\langle \xi\right.  \left\vert \xi\right\rangle }=l^{\prime}+\frac
{1}{2\Theta_{3}\left(  l^{\prime}\mid i\pi\right)  }\frac{\partial\Theta
_{3}\left(  l^{\prime}\mid i\pi\right)  }{\partial l} \tag{40}%
\end{equation}
whichs can be expanded using the following identity for the theta functions%
\begin{equation}
\frac{\partial\Theta_{3}\left(  \nu\right)  }{\partial\nu}=\pi\Theta
_{3}\left(  \nu\right)  \left(  \underset{n=1}{\overset{\infty}{\sum}}%
\frac{2iq^{2n-1}e^{2i\pi\nu}}{1+q^{2n-1}e^{2i\pi\nu}}-\underset{n=1}%
{\overset{\infty}{\sum}}\frac{2iqe^{-2i\pi\nu}}{1+q^{2n-1}e^{-2i\pi\nu}%
}\right)  \tag{41}%
\end{equation}
given explicitly%
\begin{equation}
\frac{\left\langle \xi\right\vert \widehat{J}\left\vert \xi\right\rangle
}{\left\langle \xi\right.  \left\vert \xi\right\rangle }=l^{\prime}+2\pi
Sin\left(  2l^{\prime}\pi\right)  \underset{n=1}{\overset{\infty}{\sum}}%
\frac{e^{-\pi^{2}\left(  2n-1\right)  }}{\left(  1+e^{-\pi^{2}\left(
2n-1\right)  }e^{2i\pi l^{\prime}}\right)  \left(  1+e^{-\pi^{2}\left(
2n-1\right)  }e^{-2i\pi l^{\prime}}\right)  } \tag{42}%
\end{equation}
Notice the important result coming from the above expression: the fourth
condition required for the CS demands not only $l$ to be integer or
semi-integer (as the case for the circle quantization) but also that%
\begin{equation}
\varphi=\left(  2k+1\right)  \pi\tag{43}%
\end{equation}
that leads a natural quantization similar as the charge quantization in the
Dirac monopole. Precisely this condition over the angle leads the position of
the particle in the internal or the external border of the M\"{o}bius band,
that for $r=\frac{1}{2}$ is $s=\pm\frac{1}{2}$ how is requested to be.

In order to compare our case with the CS constructed in [3] we consider the
existence of the unitary operator $U\equiv e^{i\varphi}$, such that $[J,U]=U$
then $U\left\vert j\right\rangle =\left\vert j+1\right\rangle $ such the same
average as in the previous case for the $\widehat{J}$ operator is:%
\begin{align}
\frac{\left\langle \xi\right\vert U\left\vert \xi\right\rangle }{\left\langle
\xi\right.  \left\vert \xi\right\rangle }  &  =e^{-\frac{1}{4}}e^{i\varphi
}\frac{\Theta_{2}\left(  \frac{il^{\prime}}{\pi}\mid\frac{i}{\pi}\right)
}{\Theta_{3}\left(  \frac{il^{\prime}}{\pi}\mid\frac{i}{\pi}\right)  }%
\tag{45}\\
&  =e^{-\frac{1}{4}}e^{i\varphi}\frac{\Theta_{3}\left(  l^{\prime}+1/2\mid
i\pi\right)  }{\Theta_{3}\left(  l^{\prime}\mid i\pi\right)  }\nonumber
\end{align}
where in the last equality the relation $\Theta_{2}\left(  \nu\right)
=e^{i\pi\left(  \frac{1}{4}\tau+\nu\right)  }\Theta_{3}\left(  \nu
+\tau/2\right)  $ was introduced. Also as in [3], \ we can make the relative
average for the operator $U$ in order to eliminate the factor $e^{-\frac{1}%
{4}}$, then at the first order expression (45) coincides with the unitary
circle. It is clear that the denominator in the quotient (45), average with
respect to the fiducial CS state, plays the role to centralize the expression
of the numerator. However, the claim that $U$ is the best candidate for the
position operator is still obscure and requires special analysis that we will
be given elsewhere [8].

\section{Quantum mechanics in the M\"{o}bius strip}

The Hamiltonian at quantum level operates as follows
\begin{equation}
\widehat{H}\left\vert E\right\rangle =E\left\vert E\right\rangle
\ \ \ if\ \ \ \ \left\vert E\right\rangle =\left\vert j\right\rangle
\rightarrow\tag{46}%
\end{equation}%
\begin{equation}
E=\left\{  \frac{\left(  j+\frac{rL_{0}Cos\left(  \varphi/2\right)  }%
{2}\right)  ^{2}\left[  (1+r\ Cos\left(  \varphi/2\right)  \ )^{2}-\frac
{r^{2}}{4}Cos\varphi\right]  }{\left[  (1+r\ Cos\left(  \varphi/2\right)
\ )^{2}+\frac{r^{2}}{4}Sin^{2}\left(  \varphi/2\right)  \right]  ^{2}}%
+L_{0}^{2}\right\}  \tag{47}%
\end{equation}

Imposing the fourth requirement, namely $\left\langle \widehat{J}\right\rangle
=l$ for the CS to the expressions , we have $\varphi=(2k+1)\pi$ and the
expression (47) for the energy takes the form%

\begin{equation}
E=\frac{2j^{2}}{4+r^{2}}+\frac{L_{0}^{2}}{2} \tag{48}%
\end{equation}
From the dynamical expressions given above, it is not difficult to make the
following remarks:

1) the Hamiltonian is not a priori, $T$ invariant. The $H_{MS}$ is $T$
invariant iff $TL_{0}=-L_{0}$: the variable conjugate to the external momenta
$l$ changes under $T$ as $J$ manifesting with this symmetry the full inversion
of the motion of the particle on a M\"{o}bius strip (evidently it is not the
case of the particle motion on the circle).

2) the distribution of energies is Gaussian: from the Bargmann representation
[1]%
\begin{equation}
\phi_{j}\left(  \xi^{\ast}\right)  \equiv\left\langle \xi\right\vert \left.
E\right\rangle =\left(  \xi^{\ast}\right)  ^{-j}e^{-\frac{j^{2}}{2}} \tag{49}%
\end{equation}
the distribution of energies is easily found
\begin{equation}
\frac{\left\vert \left\langle j\right.  \left\vert \xi\right\rangle
\right\vert ^{2}}{\left\langle \xi\right.  \left\vert \xi\right\rangle }%
=\frac{\left\vert \xi\right\vert ^{-2j}e^{-j^{2}}}{\Theta_{3}\left(  \frac
{i}{\pi}ln\left\vert \xi\right\vert \mid\frac{i}{\pi}\right)  }=\frac
{e^{-2l^{\prime}j}e^{-j^{2}}}{\Theta_{3}\left(  \frac{i}{\pi}l^{\prime}%
\mid\frac{i}{\pi}\right)  }. \tag{50}%
\end{equation}
By the other hand, using the approximate relation from the definition of the
Theta function
\begin{equation}
\Theta_{3}\left(  \frac{il^{\prime}}{\pi}\mid\frac{i}{\pi}\right)  =e^{\left(
l^{\prime}\right)  ^{2}}\sqrt{\pi}\left(  1+2\underset{n=1}{\overset{\infty
}{\sum}}e^{-\pi^{2}n^{2}}Cos\left(  2l^{\prime}\pi n\right)  \right)  \approx
e^{\left(  l^{\prime}\right)  ^{2}}\sqrt{\pi} \tag{51}%
\end{equation}
the expression (50)\ can be written as%
\begin{equation}
\frac{\left\vert \left\langle j\right.  \left\vert \xi\right\rangle
\right\vert ^{2}}{\left\langle \xi\right.  \left\vert \xi\right\rangle
}\approx\frac{1}{\sqrt{\pi}}e^{-\left(  j-l^{\prime}\right)  ^{2}} \tag{52}%
\end{equation}
It is useful to remark here that when $\varphi=(2k+1)\pi$ and $l=l^{\prime},$
the above equation coincides exactly in form with the boson case [3,7] but $l$
is semi-integer valued.

\section{The physical space of phase and the projection method}

In order to see how the projection method works in the context of the CS
quantization, we start from the torus as our quantum phase space. This means
that we have, previous reduction to the physical phase space via suitable
projection operators, $2n$ operators: $\theta,\overset{.}{\theta},\varphi$ and
$\overset{.}{\varphi}$.
\begin{equation}%
\begin{tabular}
[c]{l}%
$X=R\ Cos\varphi+r\ Sin\theta\ Cos\varphi$\\
$Y=R\ Sin\varphi+r\ Sin\theta\ Sin\varphi$\\
$Z=l+r\ Cos\theta$%
\end{tabular}
\ \ \ \ \ \tag{53}%
\end{equation}%
\begin{equation}%
\begin{tabular}
[c]{l}%
$\overset{.}{X}=-\overset{.}{\varphi}\ Sin\varphi\left(  R+rSin\theta\right)
+r\ Cos\theta\ Cos\varphi\overset{.}{\theta}$\\
$\overset{.}{Y}=\ \ \ \overset{.}{\varphi}\ Cos\varphi\left(  R+rSin\theta
\right)  +r\ Cos\theta\ Sin\varphi\overset{.}{\theta}$\\
$\overset{.}{Z}=\ \ \ \overset{.}{Z_{0}}-r\ Sin\theta\ \overset{.}{\theta}$%
\end{tabular}
\ \ \ \ \ \ ,Z_{0}=l \tag{54}%
\end{equation}
Then%
\begin{equation}
L_{torus}=\frac{m}{2}\left\{  \overset{.}{\varphi}^{2}\left[  (R+r\ Sin\theta
\ )^{2}+\frac{r^{2}}{4}\right]  +\left(  r\overset{.}{\theta}\right)
^{2}-2r\ Sin\theta\ \overset{.}{Z}_{0}\overset{.}{\theta}+\left(  \overset
{.}{Z}_{0}\right)  ^{2}\right\}  \tag{55}%
\end{equation}
Before we move to equations of motion of the torus is interesting to notice
that inserting the geometrical constraint $(5)$ into the above expression, the
Lagrangian of the torus becomes the Lagrangian (7) for the M\"{o}bius strip.
The Hamiltonian for the torus is easily computed from the following
expressions ($m=R=1$)%
\begin{equation}
H=p_{\theta}\overset{.}{\theta}+p_{\varphi}\overset{.}{\varphi}+p_{z_{0}%
}\overset{.}{Z}_{0}-L \tag{56}%
\end{equation}%
\begin{equation}
p_{\varphi}\equiv\frac{\partial L}{\partial\overset{.}{\varphi}}=\overset
{.}{\varphi}(1+r\ Sin\theta\ )^{2}=J_{0} \tag{57}%
\end{equation}%
\begin{equation}
p_{z_{0}}\equiv\frac{\partial L}{\partial\overset{.}{Z}_{0}}=-r\ Sin\theta
\overset{.}{\theta}+\overset{.}{Z}_{0}=L_{0} \tag{58}%
\end{equation}%
\begin{equation}
p_{\theta}\equiv\frac{\partial L}{\partial\overset{.}{\theta}}=r^{2}%
\overset{.}{\theta}-r\ Sin\theta\ \overset{.}{Z}_{0} \tag{59}%
\end{equation}%
\begin{align}
H  &  =L_{torus}=\frac{1}{2}\left\{  \overset{.}{\varphi}^{2}\left[
(1+r\ Sin\theta\ )^{2}+\frac{r^{2}}{4}\right]  +\left(  r\overset{.}{\theta
}\right)  ^{2}-2r\ Sin\theta\ \overset{.}{Z}_{0}\overset{.}{\theta}+\left(
\overset{.}{Z}_{0}\right)  ^{2}\right\} \tag{60}\\
&  =\frac{1}{2}\left\{  \frac{J_{0}^{2}}{(1+r\ Sin\theta\ )^{2}}+\frac{\left(
p_{\theta}+rSin\theta L_{0}\right)  ^{2}}{\left(  r\ Cos\theta\ \right)  ^{2}%
}+L_{0}^{2}\right\} \nonumber
\end{align}

Now we pass to construct the CS\ for the torus analogically that in the
previous Section for the M\"{o}bius strip, but in tis case the coordinate
$\theta$ are absolutely independent of $\varphi$. Thus, we assume two
"cylinder type " parametrizations: one for $0\leq l\leq$ $\infty$ cylinder
with angular variable $\varphi$ and the other one with finite $0\leq l_{2}%
\leq2\pi Sin^{2}\varphi(R=1)$%
\begin{equation}
\xi_{torus}=e^{-\left(  l+r\ Cos\theta\right)  +i\varphi}\left(
1+r\ Sin\theta\right)  e^{-2\pi Sin^{2}\varphi+k\theta}\ \ \ \ \ \ ,\ i^{2}%
=k^{2}=-1 \tag{61}%
\end{equation}
We call the above expression the \textit{geometrical} factorization. From the
above expression the physical decomposition for $\left\vert \xi_{torus}%
\right\rangle $ that is useful for our proposal is the following%
\begin{equation}
\left\vert \xi_{torus}\right\rangle =\overset{\infty}{\underset{j,m=-\infty
}{\sum}}\xi_{MS}^{-j}e^{-\frac{j^{2}}{2}}\xi^{-m}e^{-\frac{m^{2}}{2}%
}\left\vert j,m\right\rangle \tag{62}%
\end{equation}%
\[
\left\vert \xi_{MS}\right\rangle =\overset{\infty}{\underset{j,m=-\infty}%
{\sum}}\xi_{MS}^{-j}e^{-\frac{j^{2}}{2}}\left\vert j,0\right\rangle
\]
where we split the part corresponding on the Mobius strip of the rest of the
toroidal space of phase%
\begin{align}
\xi_{MS}  &  =e^{-\left(  l-r\ Sin\left(  \varphi/2\right)  \right)
+ln\left(  1+r\ Cos\left(  \varphi/2\right)  \right)  +i\varphi}\tag{63}\\
\xi &  =e^{-2\pi Sin^{2}\varphi-r\ \left(  Cos\theta+Sin\left(  \varphi
/2\right)  \right)  +ln\left(  \frac{1+rSin\theta}{1+r\ Cos\left(
\varphi/2\right)  }\right)  +k\theta}\nonumber
\end{align}
and a $m$ basis was consistently included. This factorization is the
\textit{physical} one.

We already have all ingredients to perform the projection from our toroidal
phase space to the physical phase space that we are interested in%
\begin{equation}
\left\langle \left\langle \xi_{MS}\mid\xi_{MS}^{\prime}\right\rangle
\right\rangle =\frac{\left\langle \xi_{torus}\left\vert \left\vert \xi
_{MS}\right\rangle \left\langle \xi_{MS}\right\vert \right\vert \xi
_{torus}^{\prime}\right\rangle }{\left\langle \xi_{torus}^{0}\left\vert
\xi_{MS}\right\rangle \left\langle \xi_{MS}\right\vert \xi_{torus}%
^{0}\right\rangle }=\overset{\infty}{\underset{j=-\infty}{\sum}}e^{\left(
l^{\prime}+h^{\prime}\right)  j}e^{-i\left(  \varphi-\psi\right)  }e^{-j^{2}}
\tag{64}%
\end{equation}
with , however, $\left\vert \xi_{torus}^{0}\right\rangle \equiv\left\vert
1_{torus}\right\rangle =\overset{\infty}{\underset{j,m=-\infty}{\sum}%
}e^{-\frac{m^{2}+j^{2}}{2}}\left\vert j,m\right\rangle $ . It is important to
note that we can proceed other time performing the projection from the
M\"{o}bius geometry to the circle straighforwardly obtaining the CS for the
Bose case. Then the procedure of projections can be sinthetized in the
following schema%
\[%
\begin{array}
[c]{ccc}%
Torus\rightarrow\text{Projection Op.}\rightarrow & M\ddot{o}%
bius\ strip(fermion) & \rightarrow\text{Projection Op.}\rightarrow
circle(boson)
\end{array}
\]

Besides the instructive standard procedure given above, where we take
advantage on the projection properties of the CS, there exists one powerful
method that is based on the universal projector operator%
\begin{equation}
E\left(  \theta-\frac{\pi+\varphi}{2}\leq\delta\right)  =\int_{-\infty
}^{\infty}d\lambda e^{-i\left\vert \theta-\frac{\pi+\varphi}{2}\right\vert
^{2}}\frac{Sin\left(  \delta^{2}\lambda\right)  }{\pi\lambda} \tag{65}%
\end{equation}
that clearly depends only on the constraints, being independent on the
specific form of the Hamiltonian or on the form that we factorize the original
"big" phase space. For example, it is well known that the CS defined in [2]
are a particular case that the CS\ defined in [6] by means of a displacement
operator. This fact is crucial in order to be consistent at the hour to define
correctly the observables of the physical system under consideration, in
particular the position operator [7,8].

\section{Concluding remarks}

In this work the coherent states (CS) for the fermions in the M\"{o}bius band
was constructed and compared with the previous works where the cylinder was
used and the spin $1/2$ was introduced by hand. Using these coherent states
particularly contructed we have explicitly shown, that an unoriented surface
that is the M\"{o}bius band is the natural phase space for fermionic fields.
This is because the symmetry properties of the band and the symmetry of the
fermions are closely related: both have the characteristic "double covering"
that makes that the symmetry invariance is $4\pi$ instead the $2\pi$ for the
bosonic case where the natural phase space is the cylinder. Also because the
Coherent states, due the double role that they have, namely, as projectors
[10] and making the connection between classical and quantum formulations [5]
are very sensibles to the geometrical framework where they was constructed,
given the best description of a given physical system. These important facts
permit, as we have shown also here, the reduction from the toroidal phase
space to the M\"{o}bius strip space of phase and lead, due the wonderful
propierties of the CS, a "Dirac-Like" quantization$.$

It will be interesting to construct coherent states in other geometries and
dimensions and to analyze the physical systems that they describe in such
cases. This is the main task of future works [8].

\section{Acknowledgements}

I am very thankful to Professors John Klauder for his advisements and
introducing me to the subject of coherent states and projection operators and
to E. C. G. Sudarshan for his interest demonstrated in this work in a private
communication. Also thanks are given to Professors A. Dorokhov and Yu.
Stepanovsky for my scientific formation. This work was partially supported by
PNPD-CNPQ\ brazilian funds.

\section{References}

[1]. J. R. Klauder and B. S. Skagerstam, \textit{Coherent States: Applications
in Physics and Mathemetical Physics} (World Sci., Singapore, 1985), J-P.
Gazeau, \textit{Coherent States in Quantum Physics}, Wiley-VCH, Berlin, 2009,
A. Perelomov, \textit{Generalized coherent states and their applications},
Springer, Berlin 1986.

[2]. A. O. Barut and L. Girardello, Commun. Math. Phys.21, 41 (1971).

[3]. K. Kowalski et al., J. Phys. \textbf{A 29}, 4149 (1996).

[4]. K. Kowalski and Rembielinski, Phys. Rev.\textbf{ A75}, 052102 (2007).

[5]. J. P. Gazeau and J. R. Klauder, J. Phys.\textbf{ A 32}, 123 (1999).

[6]. D. J. Cirilo-Lombardo and J.R.Klauder, in preparation.

[7]. J. A. Gonzalez and M. A. del Olmo, J. Phys. \textbf{A 31}, 8841 (1998).

[8] D. J .Cirilo-Lombardo, in preparation.

[9] D. J. Cirilo-Lombardo, Physics of Particles and Nuclei Letters \textbf{6},
No.\textbf{ 5}, 359 (2009).

[10]. A. Kempf and J. R. Klauder, J.Phys. \textbf{A 34},1019 (2001)

\end{document}